# $^{26}$Al-$^{26}$Mg isotopic, mineralogy, petrography of a Hibonite-Pyroxene Spherule in Allan Hills 77307 (CO3.03): Implications for the origin and evolution of these objects

Ritesh Kumar Mishra[1, 2*]

[1]Center for Isotope Cosmochemistry and Geochronology,

Astromaterials Research and Exploration Science division,

EISD-XI, NASA-Johnson Space Center,

 2101, NASA Parkway, Houston, TX 77058, USA,

[2]Institut fur Geoweissenchaften, Im Neuenheimer Feld 234-236

Ruprecht-Karls-Universität, Heidelberg, D 69210 Germany.

Now [*]Independent researcher, Dhawalpur Bihar 813211 India.

email: riteshkumarmishra@gmail.com

Ph:+91-766-722-7644

**Abstract**

10 Hibonite-pyroxene/glass spherules discovered hitherto are a rare suite of refractory inclusions that show the largest range of "exotic" isotopic properties (anomalies in neutron rich isotopes (e.g. $^{48}$Ca, $^{50}$Ti), abundance of $^{26}$Al) despite their defining simple spherical morphology and mineralogy consisting predominantly of few hibonites nestled within/with glassy or crystallised calcium, aluminium-rich pyroxene. $^{26}$Al-$^{26}$Mg chronological studies along with petrography and mineralogy of a relatively large (~120μm diameter), found in Allan Hills 77307 (CO3.03) has been performed. Uniquely, both hibonite and pyroxene show discordant abundance of short-lived now-extinct radionuclide $^{26}$Al that suggest disparate and distinct regions of origin of hibonite and pyroxene. The pristine petrography and mineralogy of this inclusion allow discernment of their genesis and trend of alteration in hibonite-pyroxene/glass spherules.

# 1.0 Introduction:

## 1.1 Early Solar System Evolution Records in Meteorites and Hibonite-Pyroxene Spherules

Records of the earliest events and processes during the birth and early evolution of the Solar system are preserved in various components of meteorites. The first forming solids of the Solar system- so called, Calcium, -Aluminium-rich inclusions (CAI), date the formation of the Solar system at ~ (4567.2 ± 0.5) Ma (Amelin et al., 2010; Bouvier and Wadhwa 2010). In an oxidizing environment (C/O <1) of a gas of solar composition at a total pressure of <$10^{-3}$ bar, model thermodynamical equilibrium condensation calculations predict a typical, at times overlapping, formation of corundum ($Al_2O_3$), hibonite ($CaAl_{12}O_{19}$), grossite ($CaAl_4O_7$), perovskite ($CaTiO_3$), melilite (gehlenite $CaAl_2SiO_7$-$CaMg_2Si_2O_7$ akermanite), spinel ($MgAl_2O_4$), Ca-pyroxene, feldspar, olivine (($Fe,Mg)_2SiO_4$) (Yoneda and Grossman 1995; Ebel 2006). But rapidly changing cosmochemical conditions and diverse petrogenetic setting at times only allow a transient dynamic equilibrium which plausibly is reflected in only a distinctive combination of these minerals in CAIs within different chondritic groups and types (MacPherson, 2014; Krot et al., 2014; Scott and Krot 2014). Hence, there is preponderance of some (e.g. perovskite, spinel, melilite) while paucity of others (corundum, grossite ($CaAl_4O_7$), krotite ($CaAl_2O_4$)) minerals. It is important to emphasis that modal abundances, mean sizes, and the typical mineralogies of refractory inclusions vary significantly even amongst different groups of carbonaceous chondrite (e.g. CV, CR, CO) and provide very useful information to constrain critical parameters of the prevailing cosmochemical environment such as pathways, duration, timescales, cooling rates, fugacity etc..

## 1.2 Rare abundance of hibonite bearing CAIs and hibonite- silicate inclusions

In an oxidizing environment the earliest high temperature cosmochemical conditions records are proffered by hibonite's ($CaAl_{12}O_{19}$) petrography and isotopic records since it is the 2$^{nd}$ major element mineral phase to condense at equilibrium following corundum ($Al_2O_3$). Therefore, petrographic and isotopic records of hibonite bearing refractory inclusions hold significance compared to the quintessential melilite bearing (Type B/A) CAIs. Hibonites occur in a few petrogenetic settings almost exclusively in carbonaceous chondrite either as isolated grains or in association with spinel, perovskite, and melilite or in the Wark-lovering rims of CAIs. Very rarely and obscurely in a distinct class of objects called hibonite-pyroxene spherules, a few hibonite laths are surrounded by crystalline or glassy silica-rich phase that has Ca,-Al-rich pyroxene like composition (Ireland et al., 1991). **HI**bonite-pyro**X**ene/ glass **S**pherule (hereafter referred as **HIXs**) are called so because of distinctive presence of hibonite laths, positioned mostly centrally, and surrounded spherically by pyroxene or glass of composition that is enriched in Ca, Al. Since the first descriptive work on these objects by Ireland et al., (1991) only three such studies (elemental, isotopic), for a total inventory of 10 HIXs, have been reported in the literature (Russell et al., 1998; Simon et al., 1998; Guan et al., 2000). The importance of these HIXs for understanding the formation and early evolution of the Solar system emanates from their unique exotic characteristics: (1) large anomalies in $\delta^{48}Ca$ (-45 to +45‰) and $\delta^{50}Ti$ (-60 to +105‰) (2) rare earth elements (REE) 'type II' abundance patterns in hibonite and surrounding pyroxene/glass (3) hibonites and pyroxene/glass within all HIXs (except for pyroxene in Y-791717) are devoid of 'typical' explicit (>2σ) radiogenic excess in $\Delta^{26}Mg$ (4) most show well resolved deficits in $\delta^{25}Mg$ (Kurat, 1975; Grossman et al., 1988; Ireland, 1988; Ireland et al., 1990, 1991; Tomeoka et al., 1992; Russell et al., 1998; Simon et al., 1998; Guan et al., 2000). The petrographic, elemental, and isotopic evidences taken together suggest that HIXs formation record one of the earliest, high temperature events of the rapidly evolving cosmochemical environment prior to the end of complete homogenization recorded by Type B CAIs (Mishra and Chaussidon 2014a; Chaussidon and Liu 2015). Table 1 lists some features of these ten hibonite-pyroxene/glass spherules reported so far (Fig. S1). In this paper we report $^{26}$Al-$^{26}$Mg isotope systematics studies along with petrography and mineralogy of a hibonite-pyroxene spherule in Allan Hills (ALHA) 77307 (CO3.03) to further our understanding of the formation and evolution of these exotic objects and by inference understand early events and processes in the Solar system.

**2. Meteorite and Measurement Techniques**

**2.1 Analytical Methods: Imaging and elemental abundances:** Back-scattered electron images, X-ray elemental abundances and quantitative analyses were obtained using a field-emission electron microprobe analyser JEOL 8530F hyperprobe at NASA Johnson Space Center in Houston, USA. An electron beam accelerated at 20kV, focused to a spot size of 1 micron and carrying a current intensity of 40nA was rastered over the meteoritic sample to obtain characteristic X rays. X-ray intensities and elemental abundances were obtained using a ThermoElectron ultradry SDD (silicon drift detector) energy dispersive spectrometer. Quantitative analyses were performed using a 20nA, focused (1μm) electron beam in wavelength dispersive mode. Five available spectrometers were calibrated using appropriate standards and analogous terrestrial minerals hibonite, plagioclase, sitkhin anorthite were measured following standard procedures prior to the analyses of mineral phases in HIXs. Matrix correction ($\Phi\rho Z$) scheme was applied to quantify the abundances.

**2.2 $^{26}$Al-$^{26}$Mg Isotopic analysis:** Mg and Al isotopes for $^{26}$Al-$^{26}$Mg relative chronology were carried out in multicollection mode using nanoSIMS 50 at Physical Research Laboratory, Ahmedabad, India. The analytical protocol involved using an O$^-$ primary beam of 20pA to sputter the standard and sample areas of 2-3 microns to produce secondary ions of $^{24}$Mg, $^{25}$Mg, $^{26}$Mg, $^{27}$Al that were measured simultaneously using electron multipliers in multicollection mode. HMR of >8500 was used to resolve all the major isobaric and hydride interferences. Madagascar hibonite, Burma spinel and MORB analogous terrestrial standards were analyzed to estimate instrumental mass fractionation and yield. ROIs of homogenous regions were selected from the 50-100 sequential ion images obtained during each analysis/run. Typical analysis time 8-12 secs for each frame was set. NMR was used for magnetic stability and high mass resolution scan was obtained every 15-25 frames to automatically center the counting of ions at the centre of the flat top peaks of the high mass resolution scans. EOS and secondary ion beam centreing protocol was run after pre-sputtering and before the start of the measurements. Mass fractionation corrected excess in $^{26}$Mg in meteorites were calculated using the equation

$\delta^i Mg = [(^i Mg/^{24}Mg)_{measured}/(^i Mg/^{24}Mg)_{standard} -1]*1000$; i= 25, 26

$\Delta^{26}Mg = \delta^{26}Mg - [(1+ \delta^{25}Mg/1000)^\beta -1]*1000$

where $^{25}Mg/^{24}Mg = 0.12663$, $^{26}Mg/^{24}Mg = 0.13932$ (Catanzaro et al., 1986), $\beta = 0.5128$ (Davis et al., 2015). Weighted average errors of standards were added in quadrature to standard deviations of measurements in the meteoritic sample to obtain the reported errors.

## 3. Results

### 3.1 Petrography

Previous petrographic, isotopic studies of different components of ALHA77307 have enabled its classification to petrographic grade 3.03 (weathering grade Ae) (Bonal et al., 2007; Kimura et al., 2008) that has seen peak metamorphic temperature of ~400°C (Huss et al., 2004). National Institute of Polar Research, Tokyo kindly provided a thin section of the ALHA77307 meteorite for the study.

The hibonite-pyroxene spherule found in the ALHA77307(hereafter referred as RM-1) is a spherical object with exposed radial diameter of about ~120μm (Fig. 1). It has four fragmented hibonite laths that are positioned centrally within the object. Two large hibonite laths of ~30×15μm are surrounded by Ca, Al-rich pyroxene with wedged shaped slivers having sub-micron sized grains of Fe-,Ni metal, and FeS subsumed in an unidentified Mg-rich phase. The slivers are located towards the two opposite sides of the outer rims. The hibonite shows pale blue pleochroism while the pyroxene appears colourless with nearly uniform extinction. The two fragmental pieces of hibonites near the two other larger central laths fit the broken edges of the larger hibonite laths (Fig. 1, Fig. S2 see dotted lines around the fragmented portion 'now' and the inferred 'original' position). The edges of the euhedral/subhedral hibonite prismatic crystals along the longer axis (c-axis) are sharp while the two fragmented smaller hibonites seem to be the chipped/broken off part of the larger laths. A reconstruction of hibonite laths (for HIXs in E4631-3 also) are facilitated by the prismatic crystal structure and presence of only two hibonites. Presence of several anhedral hibonites with round edges, do not allow such explicit reconstruction for any other HIXs but are redolent. However, along the shorter axis on the lower side of the hibonite laths the sharp cleavage planes are absent and (Fig. 1H) this feature is made ambiguous. Two larger HIXs have been previously reported in Colony (SP1 ~170μm) and ALH 82101 (SP15 ~140μm) (Russell et al., 1998). Hibonite laths in HIXs within Lancé, E4631-3 show remarkable morphological similarity in size, cleavage plane, occurrence within the object with ALHA77307 compared to others HIXs in Murray and Murchison where there are presence of several fragmented, jagged hibonite scattered randomly within the glass. The Ca, Al-rich pyroxene like composition material surrounding the hibonites in HIXs, span textures ranging from glassy (Lancé 3413-1/31, MUR 7-228) to devitrified glass (Colony SP1, SP15; E4631-3) to crystallized pyroxene grains (MYSM3, Y17-6). The outer boundary of the RM-1 has smooth circular contact (Fig. 1A) quite similar to the Lancé (Fig. S1). The HIXs is separated from the matrix by a circular fracture around the object. It also has a few cracks within the pyroxene and hibonite that terminate within the object. The circular fracture and cracks are likely to be the result of thin section preparation as the surrounding loosely bound fine- grained matrix material was lost. Fine-grained fayalitic matrix material along with fragmented olivine, pyroxene, Fe,-Ni metal, sulphides grains and other CAIs are present in the close proximity to the spherule (Fig. 1B). There is greater abundance of matrix material in an (exposed 2D) area of ~250μm around HIXs in ALHA77307 compared to other areas in the thin section. There exist only one additional such 'matrix rich' pocket in the thin section. Most CAIs and chondrules in ALHA77307 have 'accretionary rims' of sizes 20-150μm that often also have a continuous outer rim of Fe-rich metal or sulphide grains. Such an 'accretionary' rim around HIXs in ALHA77307 is conspicuous.

**3.2 Mineralogy**

**3.2.1 Hibonite: Homogeneous, Fragments**

The X-ray (Kα) elemental maps of Al, Mg, Ca, Ti, Si are shown in Fig. 1. C-G. The elemental distributions within the object display sharp boundaries. For example, Al, Ca, Si maps show sharp delineation between hibonite laths and surrounding pyroxene.

The hibonites have uniform composition with abundance of $Al_2O_3$, MgO, CaO, and $TiO_2$ of 89.5, 0.8 (±0.1), 8.5(±0.1), and 1.8 (±0.1) wt. %, respectively (Fig.1, 2, 3 Table 2a) and is similar to other HIXs. The hibonite compositions ($Al_2O_3$ 88-90%, CaO ~8.5%, $TiO_2$ 1.5-2.5%) within different HIXs are quite similar and are independent of the composition of their surrounding glassy phase or Ca,-Al-rich pyroxene. The deviation from ideal composition of hibonite of $CaAl_{12}O_{19}$ is commonly observed in meteorites because of coupled substitution of $Mg^{+2}$ and $Ti^{4+}$ for $2Al^{3+}$. Theoretically, titanium concentration ($TiO_2$) up to ~8 wt.% can be incorporated in the crystal structure of hibonite (Allen et al., 1978) and has been observed in CAIs in various groups of carbonaceous chondrite (MacPherson, 2014). However, the titanium ($TiO_2$) abundances within hibonite laths in HIXs show a limited range of 1-2.5 wt.%, in particular in RM-1it is <2.0 wt.%. During the quantification, titanium was assumed to be in the quadrivalent state although some of it may also be present as trivalent charged state (Fig. 4). Presence of about ~12% of total titanium cations as trivalent cation can explain the observed excess in (Si+Ti) cation and the expected 1:1 correspondence with Mg cation. Spinel and perovskite are ubiquitous in CAIs and found abundantly in various petrogenetic settings at times also associated with or within hibonites. Four sub-micron perovskite grains are hosted within hibonite in RM-1. Previously isolated perovskite grains (<2 μm) both within hibonite and pyroxene have been found within in HIXs from Murchison 7-753, E4631-3, and ALH 85085 (Grossman, 1988; Ireland et al., 1991; Guan et al., 2000). A representative quantitative profile of elemental abundance (Fig. 1, line a, equidistance total length 20μm) in hibonite in RM-1 is shown in Fig. 2. A positive correlation between MgO (0.7 to 1.1) and $TiO_2^{tot}$ (1.5 to 1.9) abundance within hibonite in RM-1 can be clearly seen. A similar linear correlation over wider range of MgO and TiO$_2$ has been observed for Y17-6 and MYSM3 (Simon et al., 1998).

**3.2.2 Ca, Al-rich pyroxene: eponymous, crystalline, Ca-Tschermak component**

Two representative elemental abundance profiles, along two transects in the pyroxene, indicated in Fig.1A are shown in Fig. 2 (lines 'b' and 'c'). Line 'b' profiles a traverse across the entire homogeneous region to the opposite diametrical end while line 'c' follows the Mg enriched region at the boundary of the pyroxene with the sliver zone (Table 2b). The line profile 'b' shows general homogeneity (over ~100 μm) along the entire stretch. At the outer boundary regions, however a slight enrichment in Al$_2$O$_3$ (~2.0 wt. %), which is compensated by combined lower abundance of SiO$_2$ and MgO, is observed (Fig. 1, Fig. 2). The pyroxene in the central region has CaO and Al$_2$O$_3$ abundances of ~ 25 and ~30 wt. %, respectively. The expected inverse correlation between the abundances Al$_2$O$_3$ with MgO and SiO$_2$ of these resulting from crystallization of pyroxene is inferred from Fig. 2. The CaO abundance however is markedly uniform. The line profile 'c' traverses the region around the sliver zone. A significant variation in abundance of Al$_2$O$_3$, MgO and SiO$_2$ over a short distance (5-10μm; entire length of profile is ~20μm) can be noted. The Ca, Al-rich pyroxene surrounding the hibonite shows a mild zonation in Al, Ti with marginally Al-rich and enriched Ti concentration in the outer regions. The correlations between these elemental abundances within pyroxene in HIXs have been previously noted (See, Fig. 3, 4 of Simon et al., 1998). The Ca,-Al-rich pyroxene found in HIXs is primarily a solid solution of Ca-Tschermak's molecule (CaTs; CaAl$_2$Si$_2$O$_6$, Kushiroite) and diopside (Di; CaMgSi$_2$O$_6$) with a minor Titanium-pyroxene (Tpx; CaTiAl$_2$O$_6$) component. The composition in the central region corresponds to CaTs 55- 67 while the Mg rich at the edge correspond to 27-45. The observed maximum CaTs composition in RM-1 is less than those found in MYSM3 and Y 17-6.

**3.2.3 Accessory phases in the slivers: Auxillary, axillary, and ancillary**

One of the most salient features of the HIXs in ALHA77307 is the presence of unaltered phases in the slivers at the outer regions. The region identified, as sliver zone in RM-1 in ALHA77307 constitute a significantly greater part in ALHA 82101 and Y-791717. They are less prominent in the smaller Lancé inclusion (See Fig. 1 of Ireland et al., 1991). The forsterite matrix like material adhering to the HIXs in CM2 meteorites, obtained by freeze-thaw disaggregation method, could be part of such zone (Ireland et al., 1991; Russell et al., 1998; Simon et al., 1998). Several phases are present at the outer margins of the RM-1 and consist of a mélange of poorly/rapidly crystallized grains. These phases in the sliver zones are a minor component of the HIXs. However, their textures, location, and mineral compositions provide important information about the formation and alteration of HIXs. The mushy sliver zones host submicron Fe, Ni metal grain, FeS, and an unidentified Mg-, Ca-, Si bearing phase and are irregularly shaped, poorly crystallized and randomly oriented. Such a texture could form in melt inclusion kind of texture by assimilation of most of the elements incompatible to the crystal into last dreg irregular pocket. The incompatible elements making the initial bulk composition of the Al-rich, Ca-pyroxene melt were sequestered into wedged slivers/ melt inclusion kind of texture at the outer regions of this HIXs. The matrix surrounding the HIXs in ALHA77307 contains fine-grained fayalitic olivine, Fe,-Ni metal grains, FeS etc.. The mushy, irregular textured grains in the slivers are devoid of any connecting melt/vein and lack any mineralogical correlation with the surrounding matrix. These textural features and the very low petrologic type (3.03) of the meteorite suggest that these phases in the sliver zones are the representative last dregs of the melt and not alteration products. Alternatively, the mineral phases in the sliver zone could be interpreted as the trapped matrix material. This interpretation is not favored because (1) the melting of Ca-pyroxene should have also melted the matrix material and got assimilated into the bulk composition of the melt (2) similar regions of larger dimensions are observed in other HIXs.

### 3.3 $^{26}$Al-$^{26}$Mg isotopic analysis:

Isotopic analyses were carried out in both the laths of hibonites and along two perpendicular transect in the pyroxene. The isotopic analysis yielded (1) well resolved mass fractionation corrected radiogenic excesses in $^{26}$Mg of ~5 (±4.5; 2σ)‰ within hibonite (Fig. 5). (2) comparatively smaller radiogenic excesses in $^{26}$Mg of 2.5 (±2; 2σ)‰ in pyroxene (Ca-Tschermak component) in different regions within the spherule. The concomitant observed radiogenic excess in $^{26}$Mg is the distinctive feature of this spherule. (3) Both hibonite and pyroxene show similar level of fractionation F (Mg) in magnesium isotope composition ($\delta^{25}$Mg) of ~10 (±2)‰.

## 4 Discussion

**4.1 Inferences from $^{26}$Al-$^{26}$Mg isotopic studies:** The present study, unlike all previous such studies of HIXs, provides resolved radiogenic excesses in $^{26}$Mg in both hibonite and pyroxene. While hibonite analyses yield radiogenic excesses of ~ 5 ‰, the pyroxene show resolved excess of ~ 2.5 (±2; 2σ) ‰. An error weighted linear regression (Isoplot Model 1) of the entire data set gives $^{26}$Al/$^{27}$Al of (4.8±4.7)×10$^{-6}$ (95% conf.) with intercept of (2.4 ± 1.0) ‰ (Fig. 5). The obtained $^{26}$Al/$^{27}$Al isochron taken at the face value implies formation of this spherule at ~ 2.5 Ma after the formation of the homogenous solar system reservoir characterized by $^{26}$Al/$^{27}$Al of (5.25±0.15)×10$^{-5}$. Such a late formation of a hibonite bearing spherule and simultaneously high initial

$^{26}$Mg/$^{24}$Mg isotopic composition is not consistent implying either the hibonite and pyroxene did not form from a single reservoir or disturbance in the isotopic records. The petrography, elemental abundances, and mineralogy of the HIXs limit any disturbance in the isotopic records. Model isochrons (force fitted to terrestrial) for pyroxene and hibonite data separately give $^{26}$Al/$^{27}$Al values of $(9.9\pm5.2)\times10^{-5}$ and $(8.2\pm4.6)\times10^{-6}$, respectively. The supra canonical $^{26}$Al/$^{27}$Al in the pyroxene, a later forming mineral in typical condensation sequence, and an order of magnitude lower $^{26}$Al abundance in hibonite are at odds in the general scenario. Radiogenic $^{26}$Mg has been previously observed in hibonite in Murchison 7-228 ($1.5\pm1.1$; $2\sigma$) only. This radiogenic $^{26}$Mg excess in Murchison 7-228 and unresolved values in ALH 82101 (SP15 $1.3\pm3.0$; $2\sigma$) within hibonite in two HIXs (glassy spherules) when combined with depleted $^{25}$Mg in the adjoining glassy phase yielded $^{26}$Al/$^{27}$Al isochron of $(1.7\pm0.7)\times10^{-5}$ and $(4.4\pm3.4)\times10^{-6}$ for Murchison and ALH 82101, respectively. On the other hand resolved radiogenic $^{26}$Mg excess in the pyroxene has been observed in Y-791717 (17-6 spherule) 'only' which lie on the $^{26}$Al/$^{27}$Al corresponding to $(2-3)\times10^{-4}$. It is worthy to note following two points: (1) The resolved deficit in $^{25}$Mg in Murchison 7-228 and ALHA 82101 resulted in yielding a resolved isochron in them. (2) Hibonite analysis in 17-6 spherule do not yield resolved radiogenic excess in $^{26}$Mg. Other HIXs do not display any evidence of $^{26}$Al.

The Mg isotopic anomalies observed in the present study should consistently also explain other isotopic properties and anomalies observed previously. The diversity in elemental and isotopic properties of HIXs inferred in previous study can be summarized as: (1) Range of positive anomaly in $\delta^{48}$Ca from 1-40 ‰ in hibonites are usually accompanied by a slightly smaller (<10 ‰) anomalies in adjoining glass/pyroxene except for Y17-6 which has negative anomaly of ~45 ‰ in both hibonite and pyroxene.

Smaller level of anomalies in other isotopes of calcium ($^{42}$Ca, $^{43}$Ca) have been measured in a few HIXs (2) $^{50}$Ti anomalies span a larger range 1-105 ‰ in hibonite and pyroxene except for negative anomaly of -60 ‰ for Y17-6. (3) The magnitude of anomalies in both $^{48}$Ca and $^{50}$Ti within hibonites show linear correlation with the magnitude of anomalies in the surrounding pyroxene/glass phases. However, the anomalies in both the isotopes ($^{48}$Ca, $^{50}$Ti) within different HIXs do not necessarily lie on a linear trend with Y 17-6 showing the maximum deviation. The hibonites in HIXs display properties and characteristics that match partially with PLACs and SHIBs. Hibonites within HIXs show isotopic properties that range within other hibonite bearing/ rich CAIs from Murchison and Murray (carbonaceous Mighei type) meteorite. Hibonite dominated refractory inclusions in Murchison (CM type) based on the mineralogy, morphology, and isotopic properties of the refractory inclusions were classified into **PL**aty hibonite **C**rystals (PLACs), **S**pinel-**HIB**onite inclusions (SHIBs), and **B**lue **AG**gregates (BAGs) (Ireland 1988). Amongst all the solar system materials PLACs show the largest range in anomalies of $\delta^{48}$Ca (-60 to 100‰), $\delta^{50}$Ti (-70 to 280‰) and a range of $^{16}$O rich oxygen isotopic composition $\Delta^{17}$O from -28‰ to -17‰ with lower abundances of $^{26}$Al/$^{27}$Al [(1-1.5)×10$^{-5}$] (Ireland 1990; Sahijpal et al., 2000; Kööp et al., 2016a). On the hand SHIBs show uniform oxygen isotopic composition $\Delta^{17}$O of -23‰ with near canonical $^{26}$Al/$^{27}$Al abundances of ~5-1×10$^{-5}$ and much smaller anomalies in $\delta^{50}$Ti (<5‰) (Ireland 1990; Liu et al., 2009; Kööp et al., 2016b). The measured excesses in $^{26}$Mg in HIXs can be explained if a relict PLACs like hibonite with low $^{26}$Al/$^{27}$Al was incorporated in $^{26}$Al-rich pyroxene melt. This interpretation of the isotopic data is consistent with previous observations and also the proposed mechanism of formation of hibonite spherules.

**4.2 Morphological Similarity and Origin of Diversity Amongst Hibonite-Pyroxene Spherules**

Of the ten HIXs reported so far, nine have been reported in carbonaceous chondrites with five of those found in Ornans group (CO) carbonaceous chondrites of different petrologic types (3.0-3.5). These objects are characterized by their defining unique mineralogy and shape but discernable differences in accessory phase, alteration of these phases and isotopic composition exist. However, since five of these in CO type span from 3.03 to 3.5 and another three are in CM2, a discernable trend of post formation alteration and primary characteristics can be gleaned.

Classification of unequilibrated chondrites into petrologic type 3.0-3.9 is based on increasing level of thermal metamorphism experienced by them which results in characteristics shifts in several properties (e.g., texture, cathodoluminiscence of mesostases glass; abundance of presolar grains, etc.) of various components of the meteorites (Huss et al., 2004). Devitrification of glassy mesostases, reduced dispersion in $Cr_2O_3$ abundances, within chondules with increasing petrographic type due to increasing thermal metamorphism have been very well documented in chondrites. However, within the HIXs, the lowest petrographic type has crystalline pyroxene (present study, 3.03) while with the progressing petrographic type character changes to devitrified glass (Colony 3.1; ALH 82101 3.3) to glassy (Lance 3413-1/31 3.5). Therefore, the crystalline character of pyroxene or glassy characteristics of the other HIXs most probably is the result of individual cooling path the particular HIXs experienced. Slower cooling or later stage heating should typically result in formation of several micro-crystallites and/or bladed acicular Mg-rich grains in the sliver zones (See, for e.g., chondrule#1, 3 of Semarkona and chondrule#1 of Vigarano (Mishra and Goswami 2014; Mishra and Chaussidon 2012) chondrule #5 in QUE 97008 (Marhas and Mishra, 2012)). Such a texture resulting from slightly slower cooling and/or later stage reheating is seen in HIXs in ALH 82101 and Y17-6. The pyroxene in HIXs (RM-

1) in ALHA77307 shows zoning at the edges. The central (80-100μm) region around the hibonite laths is quite homogeneous and the zoning pattern in Al, Mg, Si, Ti becomes apparent towards the peripheral 20-30μm. Crystallization of pyroxene leads to zoning in elemental abundance distributions and expected relationship between $Al_2O_3$, MgO, $TiO_2$, and $SiO_2$ can be seen in Fig. 1-3 which have been also observed previously in two HIXs (CM2, MYSM3, Y17-6) (Simon et al., 1998). The qualitative constraint that can be deduced is that dust with CaTs composition were melted plausibly due to a shock or thermal event that enveloped the fragmented hibonite present proximally. The HIXs/ spherules then cooled rapidly under conditions determined by local environment (density, dust, ice?) to give the observed characteristics. Presence of thick rims around HIXs with crystalline pyroxene (present study, ALH 82101, Y-791717 Fig. 8 of Russell et al., 1998; Fig. 1 of Simon et al., 1998) are suggestive that post melting these HIXs with rims cooled relatively slowly due to trapping of heat by the surrounding dust to foster the crystalline growth. The presence of glass in HIXs from Murchison (CM2) could also have been facilitated by greater abundance of water (water/rock ratio CM 0.3-0.4 > CO 0.01-0.1) that enabled rapid quenching. The relatively slower cooling rate of RM-1 was fast enough absolutely to be consistent with (1) lack of strong crystallization zoning in the HIXs in ALHA77307 (2) presence of poorly crystallized sliver zone in several HIXs (e.g. ALHA77307, ALH 82101, Y-791717, Colony). Freeze-thaw method of obtaining HIXs from CM2 type may plausibly have led to loss of sliver zones in them because sliver zones are mostly at the outer periphery and small poorly crystallized mineral phases, porosity makes them more susceptible.

**4.3 Constraints from morphology and textural features of hibonites:**

Most of the hibonites within HIXs are shards of former larger hibonite crystals that may have broken along the weaker 'a' axis of the crystal. A first order reconstruction of the original form of hibonite is made possible in a few fortuitous, favourable scenario (see Fig. S2). The observed presence of the broken pieces of hibonites within the same HIXs implies that the breaking up of hibonites must have taken place shortly before their aggregation within the melt droplet. It is noted that not all hibonites within HIXs are broken shards. The constraint on intensity of shock is qualitative because of lack of constraints on the prevalent physical, pressure conditions. However, the shock ought to be of sufficient intensity to break a significant majority proportion of the hibonites. The shock could simultaneously also generate thermal- pressure conditions to heat, melt evaporate the ambient material. The range of minimum and maximum temperature can be inferred from the melting point of Ca-pyroxene and hibonite. But the melting point of hibonite and pyroxene depend on the total pressure, dust enrichment and hence an exact peak temperature determination is not possible. The outer edges of hibonites within HIXs in some case are rounded with anhedral edges without corresponding effects in the adjacent material. This anhedral, melted edges are indicative of incipient melting. These textural features therefore are suggestive of a short thermal pulse that induced local melting of the present dust and outer margins of some hibonites. The morphology and mineralogy of RM-1 therefore provides an end member case of the prevalent conditions. A few mineral phases that are most plausibly altered products and morphological characteristics are indicative of varying levels of modification in the original forms of HIXs. The petrography and mineral composition of hibonite and pyroxene in RM-1 are indicative of lack of any alteration to its original constitution. Thus, providing the quintessential least altered original morphological form and mineralogy constitution end member of any HIXs. The identification of the pristine form consequently enables grading all the previously studied HIXs in the order of increasing metamorphosed forms (Fig. S1). Thereby establishing the constraint for temperature- pressure- time during formation of these HIXs. A characteristic trend of increasing alteration with petrographic grade is expected amongst HIXs, but it was obfuscated by significant, most probably, parent body metamorphism of the Colony and ALHA 82101. Until the present study Colony, with petrographic grade of 3.1, was the lowest grade meteorite where HIXs has been reported. The importance of the RM-1 lies in observation/finding the pristine character that allows to discern a general trend of progressive alteration/metamorphosis of the morphological character and also possibly isotopic records.

**4.4 Evidence of alteration amongst HIXs**

Petrographic grade 3-6 indicate increasingly maximum temperature (~150 to >1000 ºC) and often longer durations of thermal events experienced by meteorites. On the lower side of the scale, petrographic grade from 3-1 is indicative of greater aqueous alteration at lower temperatures (<400ºC) with decreasing numeral indices. Therefore, HIXs within CO type of increasing petrographic grade (3.03-3.5; H4) are expected to evince increasing level of transformation, alteration amongst its components. The HIXs in ALHA77307 (CO3.03) from the present study therefore provides the end member characteristics of a HIXs with pristine characteristics. The petrographic grade of ALHA77307 (CO3.03) implies RM-1 has experienced negligible parent body metamorphism and so minerals and textures observed now represent at best the effects of nebular metamorphism to the original form. The nearly uniform mineralogy in both hibonites and pyroxene, abundances of major (e.g. Mg, Ca, Si) and trace (e.g. Na, Fe, S etc.) elements and petrography suggest absence of any modification to its primeval form. With increasing petrographic grade from 3.0 to 3.5 in CO type following observation of alteration of mineral and morphology is observed which is broadly consistent with parent body metamorphism. Veins of Fe-rich material from terrestrial weathering were found cross-cutting and surrounding the hibonite-spherule in Colony (CO3.1) (Russell et al., 1998). In ALH 82101 (CO3.3) spherule the 'sliver last-dreg zone' at the edge of the inclusion has been altered to fine-grained feldspathoid and pyroxene (Russell et al., 1998). Nepheline in a morphological similar zone was noted within the Y17-6 spherule (Simon et al., 1998) but the hibonite lath within the sliver zone was largely unaffected. Forsteritic material with several cracks was also noted in MYSM3. The analogous region in Lancé also has most probably been significantly altered to rather bear resemblance with the surrounding matrix and lose the perspective associative character with the HIXs (Fig. S1 Note the lower edge of HIXs in Lancé; Ireland et al., 1991).

**4.4.2 Alterations at the periphery and rims around HIXs:**

Parent body metamorphism is often clearly documented expectedly at the periphery and boundary of the object. The outer edges of HIXs show a clear evidence of increasing alterations, degradation of the edges of the objects with increasing petrographic type. The outer edges of HIXs in Colony, ALHA 82101, Y17-6, EET 87746, ALH 85085, and Lancé have been corroded to have uneven, jagged outer most boundary (Fig. S1). Such clearly visible distinctive marks of post formation interactions and parent body thermal metamorphism are absent in the HIXs from ALHA77307 and cannot be inferred for other HIXs which were obtained following freeze-thaw procedure. Elemental abundance maps of Na, S, Fe, do not show any obvious relationship between the HIXs (RM-1) and surrounding matrix. Thus, corroborating other evidences of the pristine nature. A similar effect of increasing alteration of the outer forsteritic rims of HIXs in ALHA77307 consisting of fine-grained silicate, metal and sulphide grains to Fe-Ni-S veins can be seen. In summary, minerals, sliver zone, circularity of the outer periphery and the rim around HIXs in ALHA77307 consistent provide evidence of its pristine character and allow to discern the trend of increasing alterations in the afore mentioned properties with the increasing petrographic type.

## 4.5 Formation Scenario of Hibonite-Pyroxene Spherules

The discordant abundance of $^{26}Al/^{27}Al$ with about two orders of magnitude lower abundance in hibonite demands the formation of hibonite from a distinct $^{26}Al$ poor reservoir than pyroxene. Simultaneously, pyroxene must have drawn its constituents from a $^{26}Al$-rich reservoir to demonstrate supra-canonical $^{26}Al/^{27}Al$. In the scenario of homogeneous early solar system such a disparity between two co-existing phases can't be explained while considering other elemental and petrographic properties indicating pristine characteristics and lack of isotopic disturbance. Thus, a reconciling scenario implies a low $^{26}Al$ abundance reservoir from which hibonite forms, being admixed with $^{26}Al$-rich injected gas and dust reservoir that forms pyroxene well before the physical and chemical homogenization of the resulting reservoir. The $\delta^{25}Mg$ ($^{25}Mg/^{24}Mg$ ratios) for both hibonite and pyroxene corresponding to positive 10‰ is suggestive of their formation from a high temperature environment where Mg loss was about 25-30%. Although the $\delta^{25}Mg$ is indistinguishable within error, the mean value of hibonite is ~12 which is marginally higher than the +9.6 observed in pyroxene. This variance is expected because of difference in condensation/ melting temperature of hibonite and pyroxene. If both hibonite and pyroxene had derived Mg from the same evaporating reservoir then by the time of formation of hibonite, pyroxene would have lost its entire inventory of Mg. So $\delta^{25}Mg$ values are also suggestive of the formation of both the mineral phases from distinct reservoirs. The average composition of hibonite, pyroxene, and the calculated bulk of the RM-1 are tabulated in Table 4. For calculating the bulk composition, hibonite and sliver zone were considered to make up about 10% and 1%, respectively while the rest of the inclusion being pyroxene. The exposed 2D surface area of hibonite to area of the HIXs is ~10% and was taken as the representative for the bulk. The calculated bulk composition of RM-1 is expectedly similar to other HIXs. The average pyroxene and calculated bulk of all HIXs are plotted on a spinel projection CMAS diagram (Fig. 6) (Huss et al., 2001; MacPherson 2014). Cooling of the calculated bulk compositions of HIXs under thermodynamical equilibrium conditions, therefore suggests formation of anorthite, melilite prior to formation of pyroxene. A pyroxene-hibonite spherule in Acfer 094 (#86 Krot et al., 2004) presents a very interesting case and possibly a connection. The spherule mineralogy consists of small regions of spinel, anorthite, melilite apart from typical hibonite and pyroxene. The object has not been formally considered as a HIXs because if spinel, anorthite and melilite are considered as constituent minerals of HIXs, a large number of other CAIs and their kindred with no common characteristic or unique features would also have to considered the part of the HIXs. However, the hibonite- pyroxene spherule in Acfer 094 (#86, Fig. 7 Krot et al., 2004) exhibits all the major unique petrographic, mineralogy, morphological characteristics of HIXs and lack of $^{26}$Al within its pyroxene and hibonite (Sugiura et al., 2007). Consideration of hibonite-pyroxene #86 in Acfer 094, is important because the meteorite (1) shows no evidence of thermal metamorphism and aqueous alteration and (2) is classified as an ungrouped type 3 carbonaceous breccia with similarities to CO-CM type. Therefore, RM-1 and Acfer 094 #86 are representative sample of HIXs with no parent body metamorphism. Interestingly, Acfer 094 hibonite-pyroxene #86 hosts the complete range of mineral phases (spinel, hibonite, anorthite, melilite, pyroxene) that would be expected to form if the bulk composition of all the HIXs were to follow equilibrium condensation path.

The rare occurrence of anorthite, melilite, spinel along with hibonite, pyroxene in Acfer 094 #86 provides a unique case to understand its thermal pathway of chemical, mineralogical evolution that is the missing in all the other HIXs. Sack and Ghiroso (2007) developed a thermodynamical model for pyroxene in chemical (mineralogy) system $CaMgSi_2O_6$ (diopside)- $CaTiAlSiO_6$ (grossmanite)- $CaMg_{1/2}Ti_{1/2}AlSiO_6$ (alumino-buffonite) – $CaAl_2SiO_6$ (kushiroite) that suggests/constraints miscibility gaps between diopside rich and alumino-buffonite -rich pyroxene (fassaite) on the diopside-alumino-buffonite join to explains the observed mineralogy in hibonite±grossite+pyroxene spherules. Thus, higher bulk titanium abundance inferred from maximum titanium abundance (~4 wt.%) in pyroxene in Acfer 094 hibonite-spherule (Krot et al., 2007) compared to other HIXs (titanium abundance $TiO_2$ <2.5 wt.%) could explain the difference. Or, alternatively Acfer 094 #86 during it formation and evolution followed /experienced a very distinct thermal path than all other HIXs in which crystallization of the mineral phases namely anorthite, and melilite were not suppressed. In summary, the petrography and mineralogy of HIXs in ALHA77307 (present study) and hibonite-pyroxene in Acfer 094 (#86) provide the pristine and end member mineralogy and composition to understand the common origin and metamorphism in HIXs. Previously, this observed inconsistency within the mineralogy of HIXs was explained by invoking rapid undercooling, the absence of nucleating sites, meta-stable states, and various rates of fast and slow cooling (0.5 °C/h- 5°C/h).

**4.6 Inferences from the Morphology and Mineralogy**

**4.6.1 Common Genesis scenario:** From a $^{26}Al$ poor early solar system reservoir hibonites form either separated temporally in an evolving reservoir or from different regions of a heterogenous reservoir simultaneously. The hibonite reservoir shares lineage with the reservoir from which PLACs and SHIBs formed. Pyroxene (Kushiroite/Calcium-Tschermak's molecule component) and akin glass formed from a $^{26}$Al-rich gas and dust reservoir that was plausibly shock heated in the approx. temperature range of 1700-1450K that allowed most of the hibonite to escape melting and re-equilibration. The melt with composition akin to CaTs trapped a few hibonites and its fragmented remains while also mixing in varying proportion. Each of the formed spherule followed a slight different cooling path/ history with crystallization of minerals/ phases determined both by the bulk initial composition and cooling rate. A relatively small difference in bulk composition of the produced melt and rate of cooling resulted in the diversity in morphology, crystalline/ glassy nature of the pyroxene, gradients in distribution of the cooling phase, formation of sliver zones. The HIXs were later incorporated in carbonaceous chondrite parent bodies and experienced an increasing level of parent body thermal and aqueous alteration constrained by the petrologic type of the parent body. The suggested origin and evolution of HIXs is quite similar to the mechanism suggested earlier by Simon et al., (1998). This work draws upon the results and observations made previously in studies of HIXs and kindreds. The suggested mechanism of formation and evolution of HIXs is able to explain the diversity in the isotopic, mineralogical and petrography of all the HIXs. HIXs, as enumerated earlier, show isotopic properties that range from PLACs to SHIBs while additionally constituted by the later forming, less refractory, silica-rich, pyroxene. Hence, despite their phylogentic and mineralogical similarities and relatively small size, these objects form a very rare suite of objects that recorded the extreme range of exotic isotopic, mineralogical properties that could help understand the early solar system event processes by forming a cog in the continuum from PLACs, SHIBs, (F)/UN CAIs to normal CAIs.

## 5. Summary and Conclusions

In one of the most pristine meteorite Allan Hills 77307 (CO3.03), a hibonite-pyroxene spherule has end member characteristics in term of morphology, mineralogy and petrography. Uniquely, both hibonite and pyroxene host well resolved but discordant amount of $^{26}Al/^{27}Al$ abundance. The isotopic study along with mineralogy and petrography suggests formation of hibonites and enveloping pyroxene from two distinct reservoirs. The end member like characteristics allow to the discern a plausible common mode of formation of the studied hibonite-pyroxene spherule and its kind. The inferred formation scenario and the increasing level of metamorphism experienced by individual hibonite- pyroxene hosted in different parent body of petrographic grades helps explain the diversity seen in these objects.

## Acknowledgements

I'm grateful to National Institute of Polar Research, Tokyo, for lending a thin section of ALHA77307. My sincerely thank Dr. D.K. Ross (Johnson Space Center, Houston) for providing generous assistance during mapping and quantitative analyses. I also acknowledge several fruitful discussions with Marc Chaussidon, Justin Simon, Kuljeet K. Marhas, Daniel K. Ross, Jangmi Han, and Alejandro Cisneros. The financial support during the work by NASA post-doctoral program (NPP) fellowship at NASA Johnson Space Center, Houston is gratefully acknowledged. The manuscript was prepared during the Humboldt fellowship at Heidelberg University. Financial support from the Alexander Von Humboldt foundation during the term is also sincerely acknowledged.

## References

Amelin Y., Kaltenbach A., Lizuka T., Stirling C. H., Ireland T. R., Petaev M., and Jacobsen S. B. U-Pb chronology of the Solar system's oldest solids with variable $^{238}U/^{235}U$. *Earth and Planetary Science Letters* **300**:343-350.

Allen J. M., Grossman L., Davis A. M., and Hutcheon I. D. 1978. Mineralogy, textures and mode of formation of a hibonite-bearing Allende inclusion. Proc. *Lunar Planet. Sci. Conf.* **IX**, 1209-1233.

Bonal L., Bourot-Denise M., Quirico E., Montagnac G., and Lewin E. 2007. Organic matter and metamorphic history of CO chondrite. *Geochimica et Cosmochimica Acta* 71:1605-1623.

Bouvier A., and Wadhwa M. 2010. The age of the Solar System redefined by the oldest Pb–Pb age of a meteoritic inclusion. *Nature Geoscience* **3**(9):637-641.

Chaussidon M., and Liu M.-C. 2015. Timing of nebula processes that shaped the precursors of the terrestrial planets. In The Early Earth: Accretion and Differentiation (eds. Badro J.) John Wiley & Sons. Pp 1-26.

Ebel D. S. 2006. condensation of rocky material in astrophysical environments. In *Meteorites and the early Solar System II* (eds. D. S. Lauretta and H. Y. McSween Jr.) Univ. Arizona Press, Tuscon. pp. 253-277.

Grossman J. N., Rubin A. E., and MacPherson G. J. 1988. ALH 85085:A unique volatile poor carbonaceous chondrite with possible implications for nebular fractionation processes. *Earth and Planetary Science Letters* **91**:33-54.

Guan Y., Huss G. R., MacPherson G. J., and Wasserburg G. J. 2000. Calcium-aluminum-rich in inclusions from enstatite chondrites:Indigenous or foreign ? *Science* **289**:1330-1333.

Huss G. R., MacPherson G. J., Wasserburg G. J., Russell S. S., and Srinivasan G. 2001. Aluminum-26 in calcium-aluminum-rich in inclusions and chondrules from unequilibrated ordinary chondrites. *Meteoritics & Planetary Science* **36**:975-997.

Huss G. R., Rubin A. E., and Grossman J. N. 2006. Thermal metamorphism in chondrites. In *Meteorites and the early Solar System II* (eds. D. S. Lauretta and H. Y. McSween Jr.) Univ. Arizona Press, Tuscon. pp. 567-586.

Ireland T. R. 1988. Correlated morphological, chemical, and isotopic characteristics of hibonites from the Murchison carbonaceous chondrite. *Geochimica et Cosmochimica Acta* **52**:2827-2839.

Ireland T. R. 1990. Presolar isotopic and chemical signatures in hibonite- bearing refractory inclusions from the Murchison carbonaceous chondrite. *Geochimica et Cosmochimica Acta* **54**:3219-3237.

Ireland T. R., Fahey A. J., and Zinner E. K. 1988. Trace-element abundances in hibonites from Murchison carbonaceous chondrite:Constraints on high temperature processes in the solar nebula. *Geochimica et Cosmochimica Acta* **52**:2841-2854.

Ireland T. R., Fahey A. J., and Zinner E. K. 1991. Hibonite-bearing microspherules: A new type of refractory inclusions with large isotopic anomalies. *Geochimica et Cosmochimica Acta* **55**:367-379.

Kimura M., Grossman J. N., and Weisberg, M. K. 2008. Fe-Ni metal in primitive chondrites:Indicators of classification and metamorphic conditions for ordinary and CO chondrites. *Meteoritics & Planetary Science* **43**:1161-1177.

Kööp L., Davis A.M., Nakashima D., Park C., Krot A. N., Nagashima K.,Tenner T. J., Heck P. R., and Kita N. T. 2016a. A link between oxygen, calcium, and titanium isotopes in $^{26}$Al-poor hibonite-rich CAIs from Murchison and implications for the heterogeneity of dust reservoirs in the solar nebula. *Geochimica et Cosmochimica Acta* **189**:70-95.

Kööp L., Nakashima D., Heck P. R., Kita N. T., Tenner T. J., Krot A. N., Nagashima K., Park C., and Davis A.M. 2016b. New constraints on the relationship between $^{26}$Al and oxygen, calcium, and titanium isotopic variation in the early Solar system from a multiple isotopic study of spinel-hibonite inclusion. *Geochimica et Cosmochimica Acta*. **184**:151-172.

Krot A. N, Keil K., Scott E. R. D., Goodrich C. A., and Weisberg M. K. Classification of meteorites and their genetic relationships. *In Davis A. M., (Ed.), Meteorites, Comets, and Planets, Treatise on Geochemistry, vol. 1* (eds. Holland H. D., and Turekian K. K.) Elsevier, 2014, pp. 1-65.

Kurat G. 1975. Der kohlige chondrit Lancé: Eine petrologishe analyse der komplexen genese eines chondriten. *Tschermaks Mineralogische Petrographische Mitteilungen* **22**:38-78.

Krot A. N., Fagan T. J., Keil K., Mckeegan K. D., Sahijpal S., Hutcheon I. D., Petaev M. I., Yurimoto H., 2004. Ca,Al-rich inclusions, amoeboid olivine aggregates, and Al-rich chondrules from the unique carbonaceous chondrite Acfer 094: I. mineralogy and petrology *Geochimica et Cosmochimica Acta*. **68**: 2167-2184.

Liu M., -C., Mckeegan K. D., Goswami J. N., Marhas K. K., Sahijpal S., Ireland T. R., and Davis A. M. 2009. Isotopic records in CM hibonites: Implications for timescales of mixing of isotope reservoirs in the solar nebula. *Geochimica et Cosmochimica Acta*. **73**: 5051-5079.

MacPherson G. J. 2014. Calcium–aluminum-rich inclusions in chondritic meteorites *In Davis A. M., (Ed.), Meteorites, Comets, and Planets, Treatise on Geochemistry, vol. 1* (eds. Holland H. D., and Turekian K. K.) Elsevier, 2014, pp. 201-246.

Marhas K. K., and Mishra R. K. 2012. Fossil Records of Fe-60 in QUE 97008 Chondrule. Meteoritics & Planetary Science, 47, A#5273 2012.

Mishra R. K., and Chaussidon M. 2014a. Timing and extent of Mg and Al isotopic homogenization in the early solar system. *Earth and Planetary Science Letters* **390**:318-326.

Mishra R. K., and Chaussidon M. 2012 $^{60}$Fe-$^{60}$Ni Isotope Systematics in Silicates in Chondrules from Unequilibrated Chondrites: Yet Again and Status Quo. Meteoritics & Planetary Science, 47, A#5194 2012.

Mishra R. K., and Goswami J. N. 2014. Fe-Ni and Al-Mg isotope records in UOC chondrules: Plausible stellar source of $^{60}$Fe and other short-lived nuclides in the early Solar system. *Geochimica et Cosmochimica Acta*. **132**:440-457.

Mishra R. K., Marhas K. K., and Sameer. 2014. Abundance of $^{60}$Fe inferred from nanoSIMS study of QUE 97008 (L3.05) chondrules. *Earth and Planetary Science Letters* **436**:71-81.

Russell S. S., Huss G. R., Fahey A. J., Greenwood R., Hutchison R. C., and Wasserburg G. J. 1998. An isotopic and petrologic study of calcium-aluminum-rich inclusions from CO3 meteorites. *Geochimica et Cosmochimica Acta*. **62**:689-714.

Sack R., Ghiorso M. S. 2017. Ti$^{3+}$ - and Ti$^{4+}$ - rich fassaites at the birth of the solar system: Thermodynamics and applications. *American Journal of Science* **317**(7):807-845.

Sahijpal S., Goswami J. N., and Davis A. M. 2000. K. Mg, Ti and Ca isotopic compositions and refractory trace element abundances in hibonites from CM and CV meteorites: implications for early solar system processes. *Geochimica et Cosmochimica Acta*. **64**:1989-2005.

Scott E. R. D., and Krot A. N. 2014. Chondrites and their components. *In Davis A.*

M., (Ed.), *Meteorites, Comets, and Planets, Treatise on Geochemistry, vol. 1* (eds. Holland H. D., and Turekian K. K.) Elsevier, 2014, pp. 66-37.

Simon S. B., Davis A. D., Grossman L., and Zinner E. K. 1998. Origin of Hibonite-pyroxene spherules found in carbonaceous chondrites. *Meteoritics & Planetary Science* **33**:411-424.

Sugiura N., and Krot A. N. 2007. $^{26}$Al-$^{26}$Mg systematics of Ca-Al-rich inclusions, amoeboid olivine aggregates, and chondrules from the ungrouped carbonaceous chondrite Acfer 094. *Meteoritics & Planetary Science* **42**:1249-1265.

Tomeoka K., Nomura K., and Takeda H. 1992. Na-bearing Ca-Al-rich inclusions in the Yamato 791717 CO carbonaceous chondrite. *Meteoritics* **27**:136-143.

Yoneda S., and Grossman L. 1995. Condensation of $CaO-MgO-Al_2O_3-SiO_2$ liquids from cosmic gases. *Geochimica et Cosmochimica Acta* **59**:3413-3444.

Table captions:

Tables:

Table 1.

General features of the hibonite-pyroxene spherules

Table 2a.

Electron microprobe analyses of hibonite

Table 2b.

Electron microprobe analyses of pyroxene

Table 3.

$^{26}$Al-$^{26}$Mg isotope data of hibonite-pyroxene spherule (RM-1) in ALHA77307 (CO3.03)

Table 4.

Chemical compositions of hibonite, pyroxene, and calculated bulk of hibonite-pyroxene spherules

Figure captions:

Figures:

Fig. 1

Back-scattered electron image (A), X-ray elemental abundance maps of Al, Mg, Ca, Ti, Si (C-G) and (B) a composite map (Mg-Ca-Al as RGB) of the hibonite-pyroxene spherule in Allan Hills 77307 (CO3.03). (H), (I) are back-scattered electron image of the 'sliver zone' and hibonite laths. The scale bar is shown in each BSE image.

Fig. 2

Representative EPMA quantitative linear profiles, indicated in Fig. 1A, across a hibonite and two transects along the surrounding pyroxene is shown. Note the analyses points along the three linear profiles (a-c) are spaced unequally on the HIXs. The linear profile 'a' ends closer to the edge of hibonite crystals and excitation volume of surrounding pyroxene is attributed for the marginal lower abundance of $Al_2O_3$ in the last 2 data points of the profile 'a'.

Fig. 3

$SiO_2$ correlation diagram in pyroxene of hibonite-pyroxene spherules

The plot shows variation of MgO, $Al_2O_3$, $TiO_2$ (Y axis) with respect to $SiO_2$ (X axis) for analyses within pyroxene. Corresponding ranges observed in pyroxene in Yamato 791717 (CO3.03) and Murray MYSM3 (CM2) (Simon et al., 1998) are shown by light and dark gray filled rectangular boxes, respectively. Open rhombuses show $Al_2O_3$ (upper) and MgO composition of pyroxene glass in Lancé, with respect to $SiO_2$ (Ireland et al., 1991).

Fig. 4

Quantitative analyses of hibonite laths from the ALHA77307 HIXs and the corresponding relationship between abundance of Mg and (Si+ Ti) cations in hibonites.

Fig. 5

$^{26}$Al isochron of hibonite-pyroxene spherule (RM-1) in ALHA77307 CO3.03. Analyses within hibonite and pyroxene are shown by circles and squares, respectively. The closed symbols indicated $\Delta^{26}Mg$ while open symbols show $\delta^{25}Mg$. The light gray and dark black solid lines correspond to $^{26}Al/^{27}Al$ of $5.25\times10^{-5}$ (Solar system canonical abundance) and $1.0\times10^{-5}$, respectively. The error weighted regression line (Isoplot model 1) $(4.8\pm4.7)\times10^{-6}$ obtained by considering all the data from RM-1 is shown by thick black dash dot dot line. Errors are 2σ.

Fig. 6

Spinel projection CMAS diagram. The bulk composition of the analysed hibonite-pyroxene spherule from ALHA77307 along with previously studied HIXs, is projected from spinel ($MgAl_2O_4$) onto the plane $Al_2O_3$-$Mg_2SiO_4$-$Ca_2SiO_4$ following Huss et al., (2001) and MacPherson (2014). Black lines show experimental phase boundaries. Condensation trends of solid phases from a gas of Solar composition are shown by the red lines. Abbreviations are Cor for corundum, Hib for hibonite, An for anorthite, Ak for akermanite, Gro for grossite, Geh for gehlenite, Di for diopside, En for enstatite, La for larnite, Mw for merwinite, Mo for monticellite, Fo for forsterite.

Fig. S1 Back scattered electron images of Hibonite-Pyroxene Spherules.

HIXs found in (A) Acfer 094 ung 3.0 (B) ALHA77307 (CO3.03) in the present study (C) in Colony (CO3.1) and (D) ALH 82101 (CO3.3) from Russell et al., 1998 (E) in Yamato 791717 (CO3.3) and (J) in Murray (CM2) from Simon et al., 1998 (F) in Lancé (CO3.5) from Kurat, 1975 (G) and (H) in ALH 85085 (CH3) from Grossman, 1988 (I) in EET 87746 from Guan et al., 2000 and (K), (L) in Murchison from Ireland et al., 1991. Scale bars are shown in each panel. B-F are hibonite-pyroxene spherule in carbonaceous Ornans type arranged in increasing petrologic type. Images have been taken from published literature. Note the trend of corrugation of the outermost boundaries, sliver zone, and alteration features in the hibonite-pyroxene spherules. See text for additional details.

Fig. S2

Inferred original structure of hibonites in Hibonite-Pyroxene Spherules

in (A) ALHA77307 (this study) (B) EET 87746 (Guan et al., 2000).

Thicker dotted lines show the positioning of the fragmental pieces (A, B, C) of hibonites prior to breaking up. Thinner lines are drawn around the fragmented pieces (A`, B`, C`) of hibonites. Some space has been deliberately left between the fragmented pieces to clearly show the edges of both the pieces. Arrows indicate the direction of motion. A dust particle was sitting at the edge of the hibonite grain in the X-ray elemental abundances maps shown in Fig. 2. It is absent in the image here. The smooth outer margin of HIXs and absence of petrographic correlations between the grains in the sliver zones and the grains in the outer matrix can be seen clearly.

Fig. 1

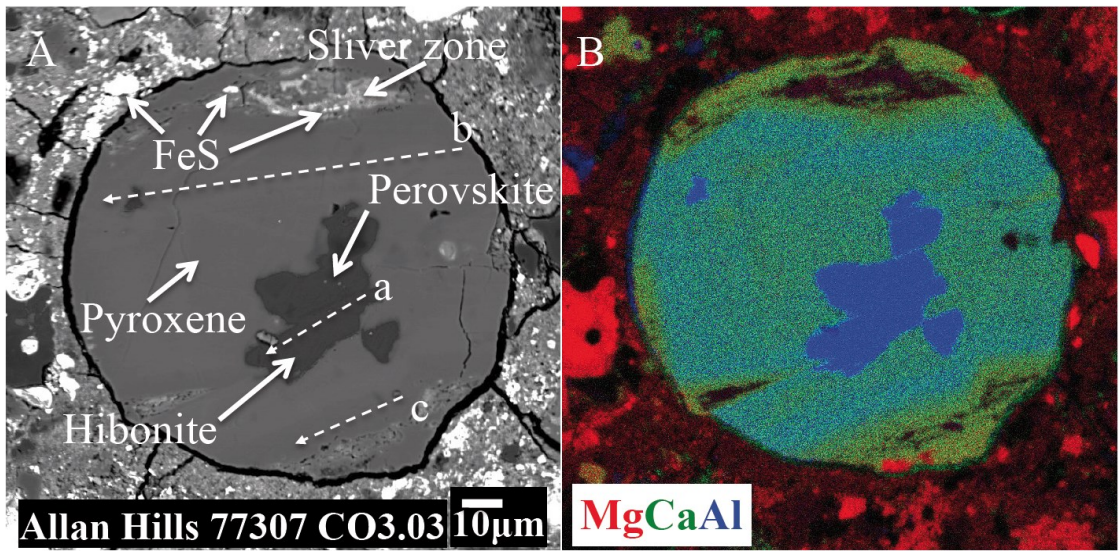

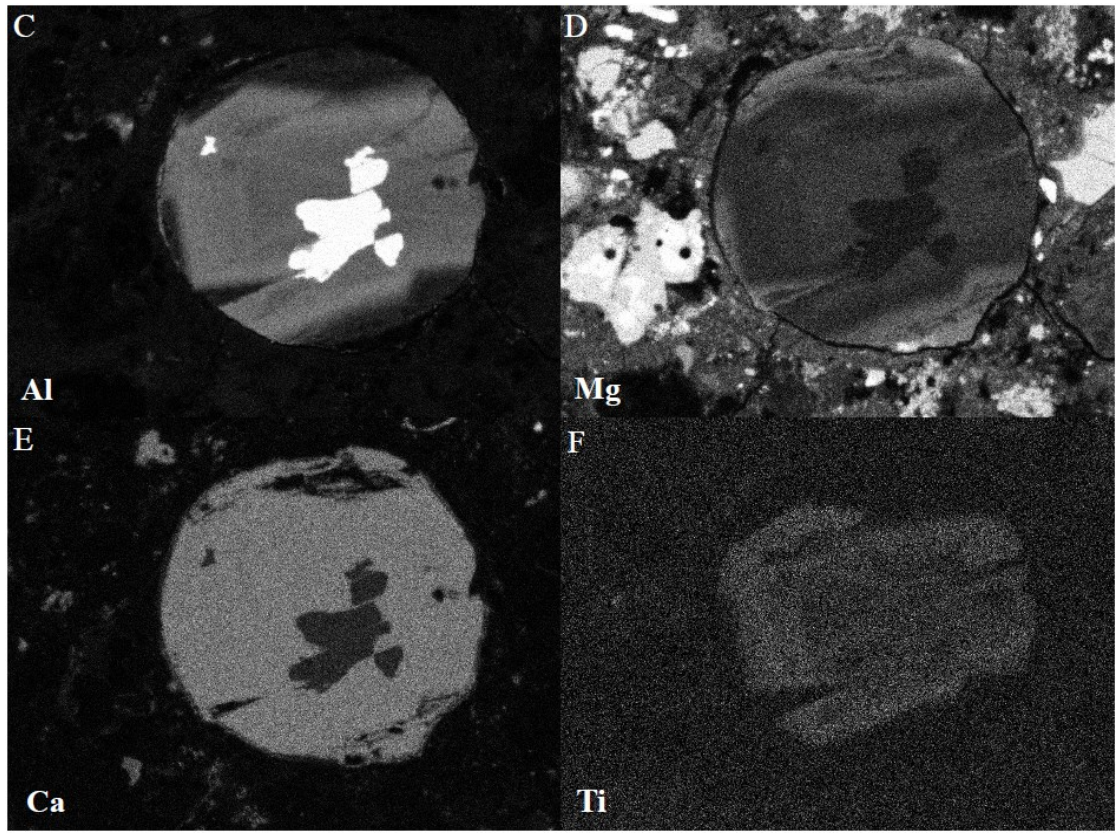
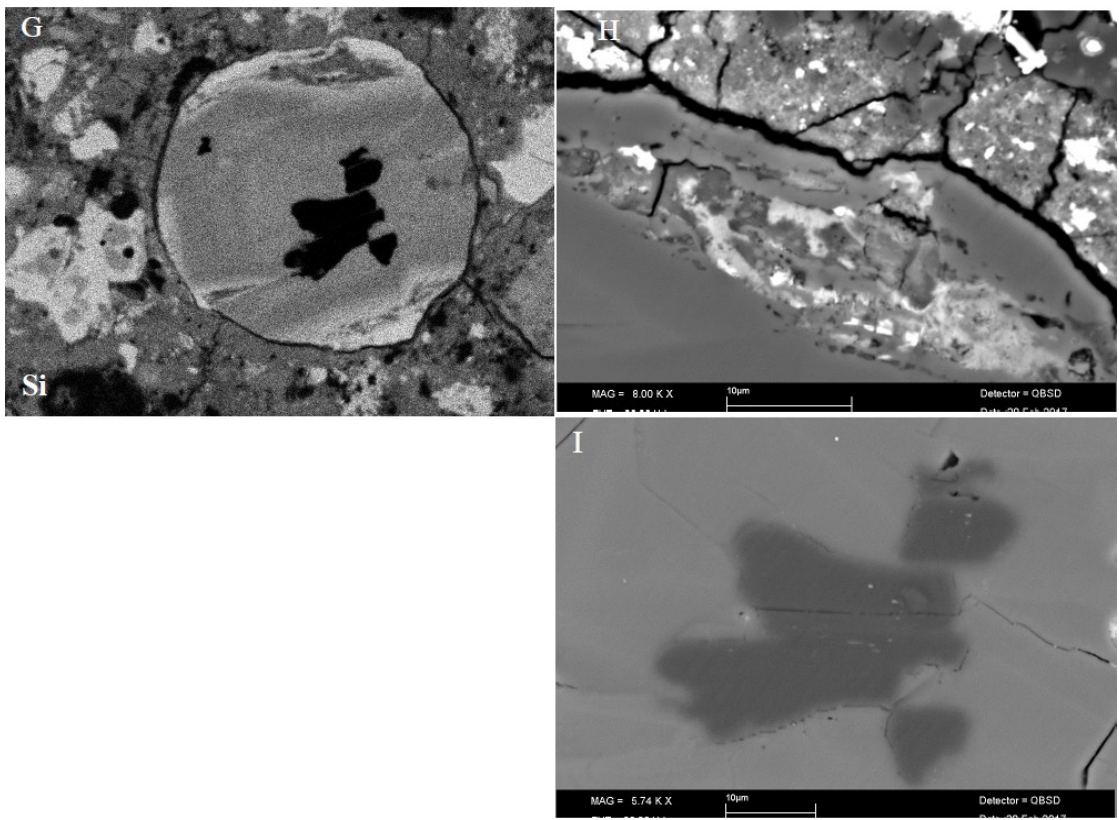

Fig. 2

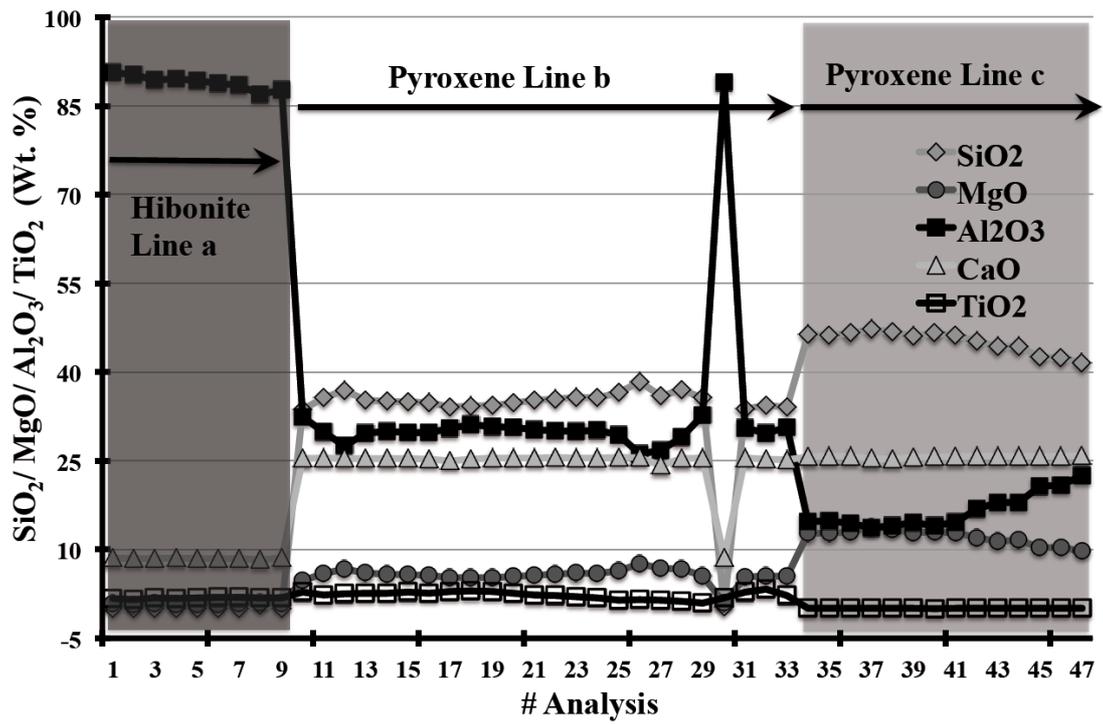

Fig. 3

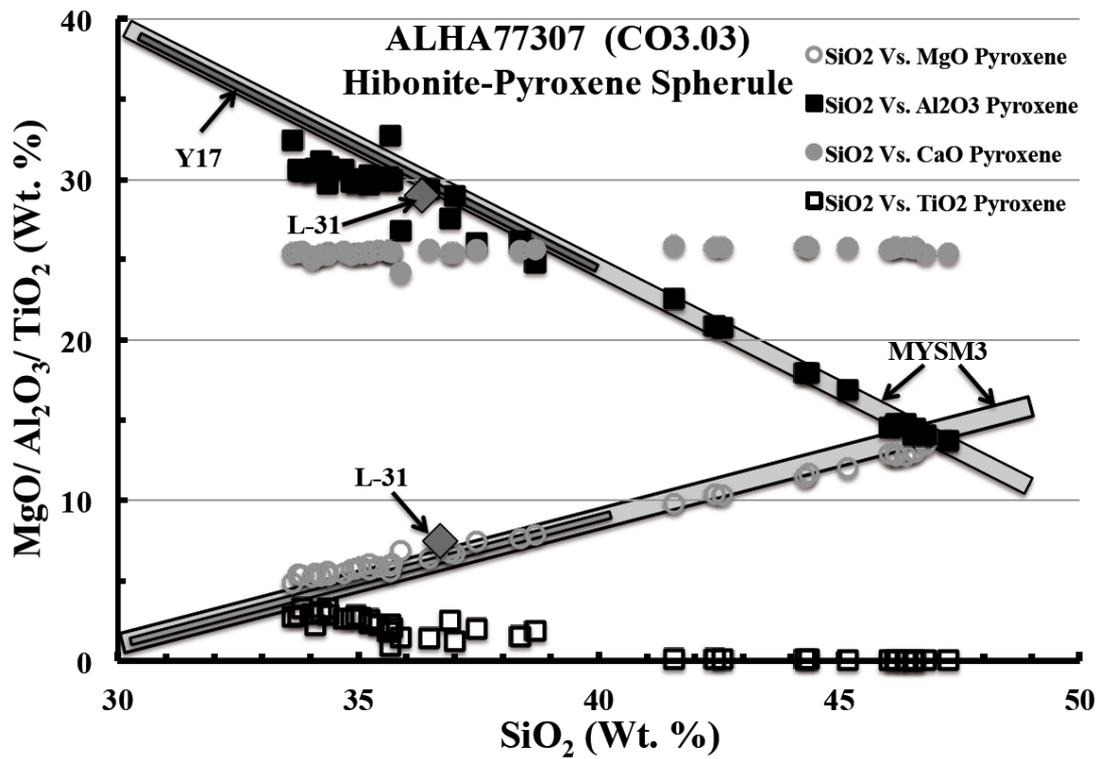

Fig. 4

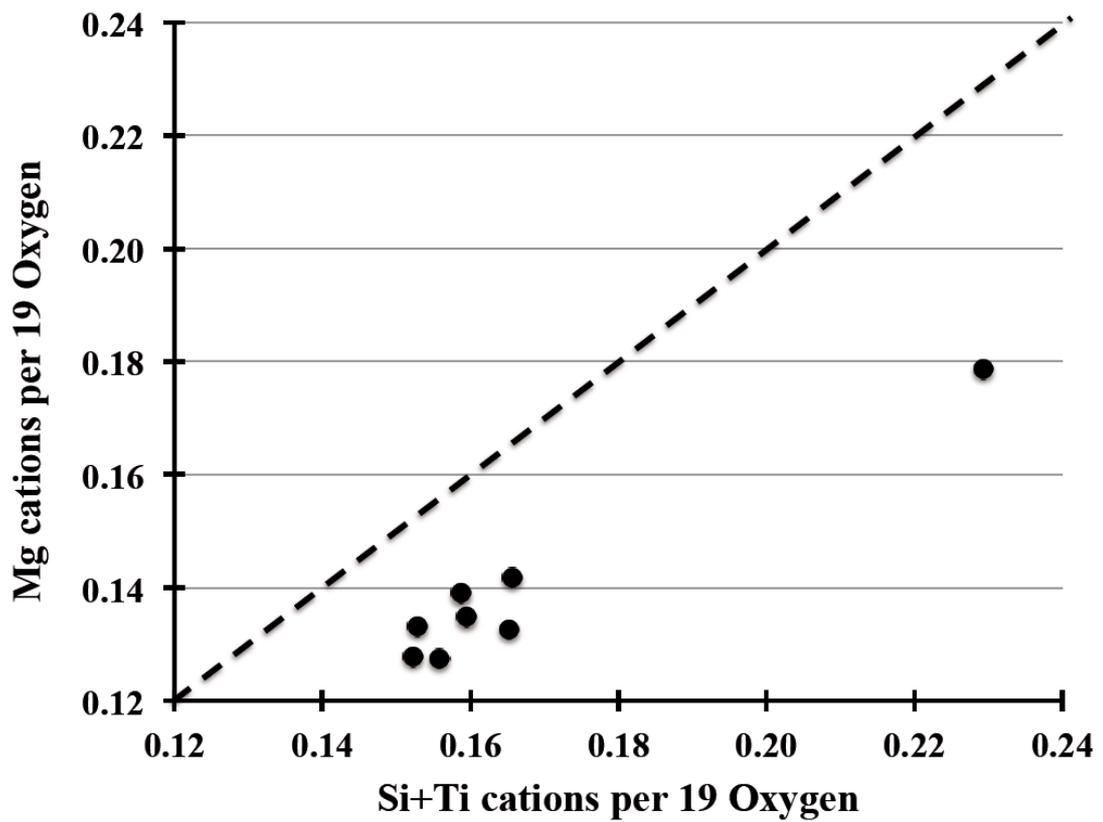

Fig. 5

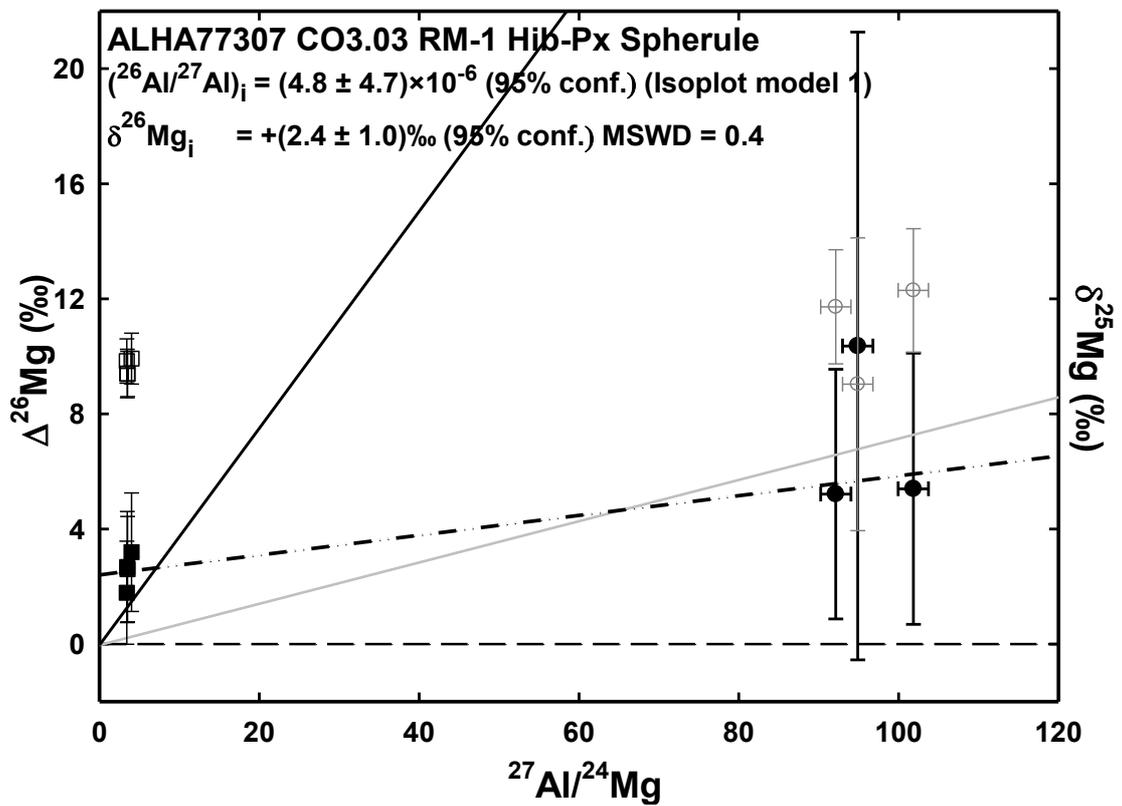

Fig. 6

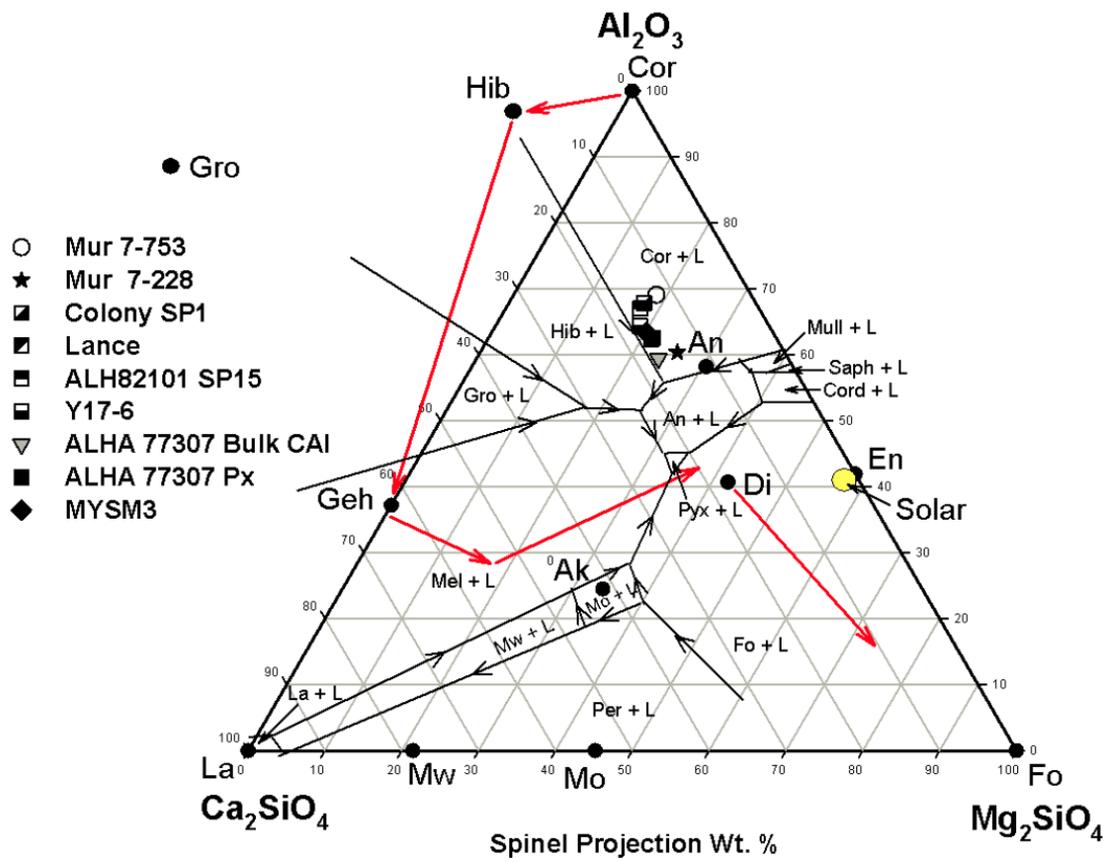

Fig. S1

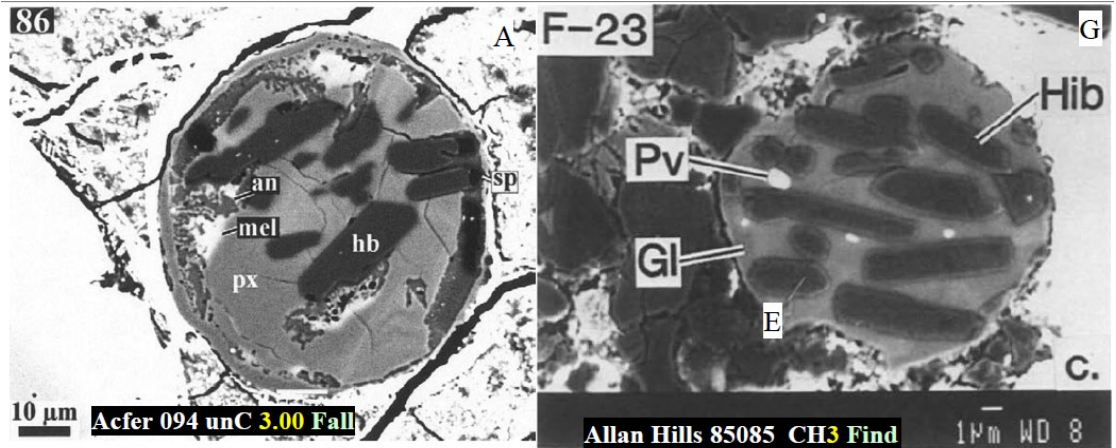
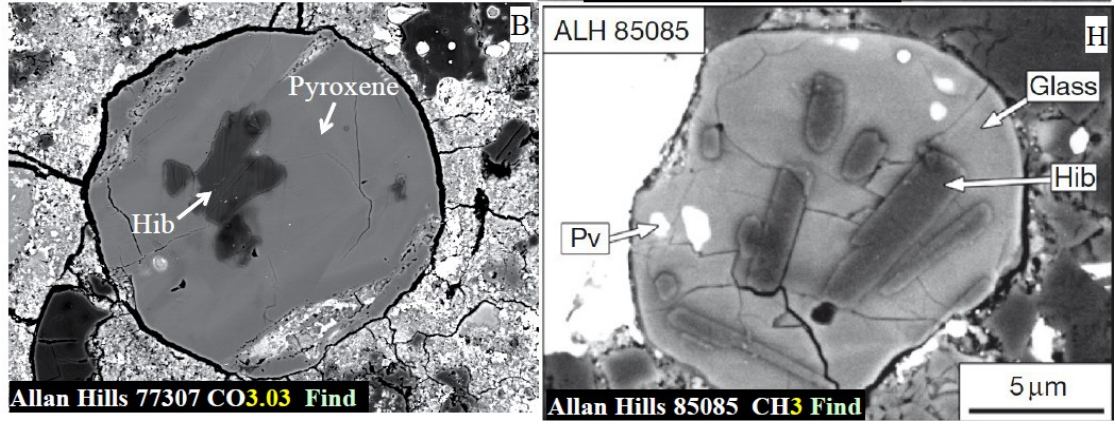
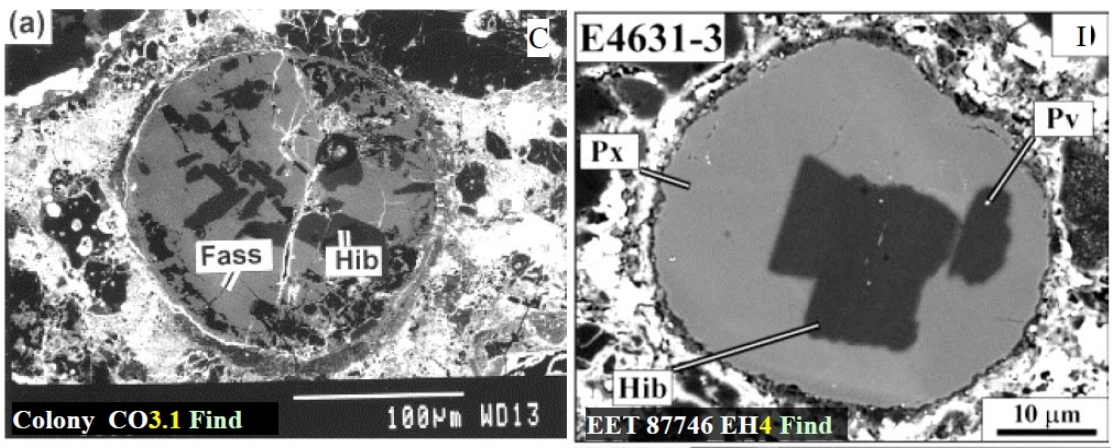
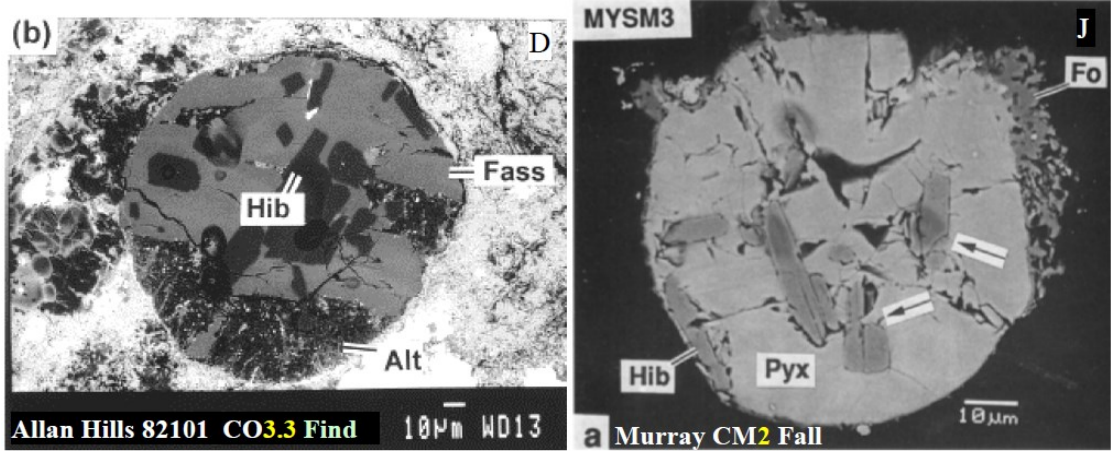

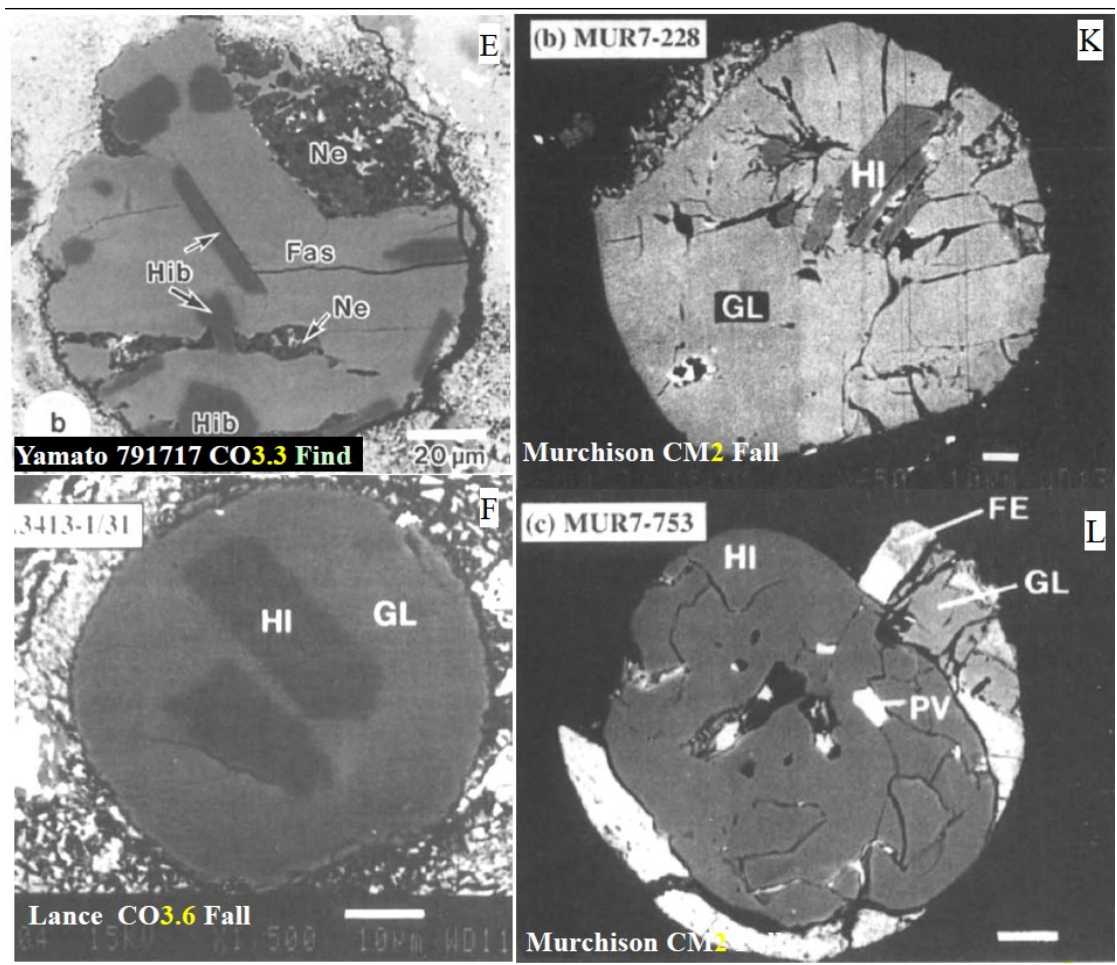

Fig. S2

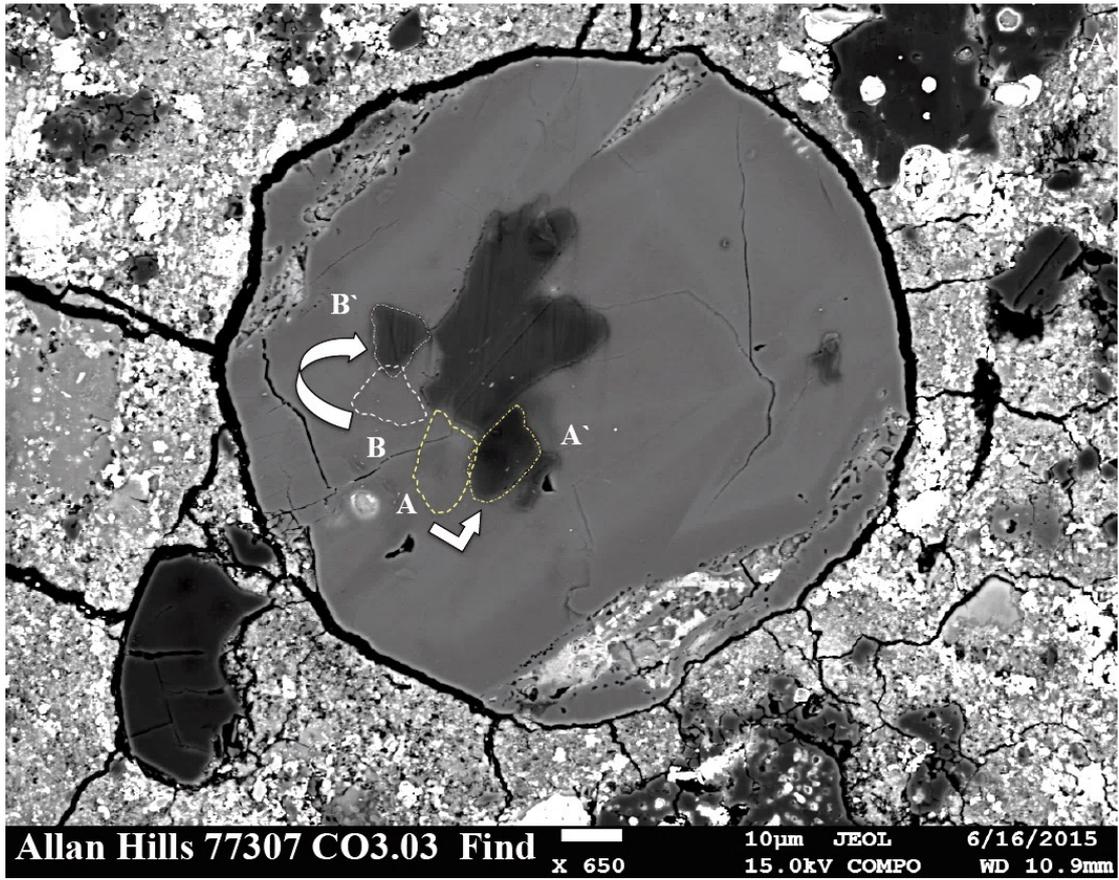

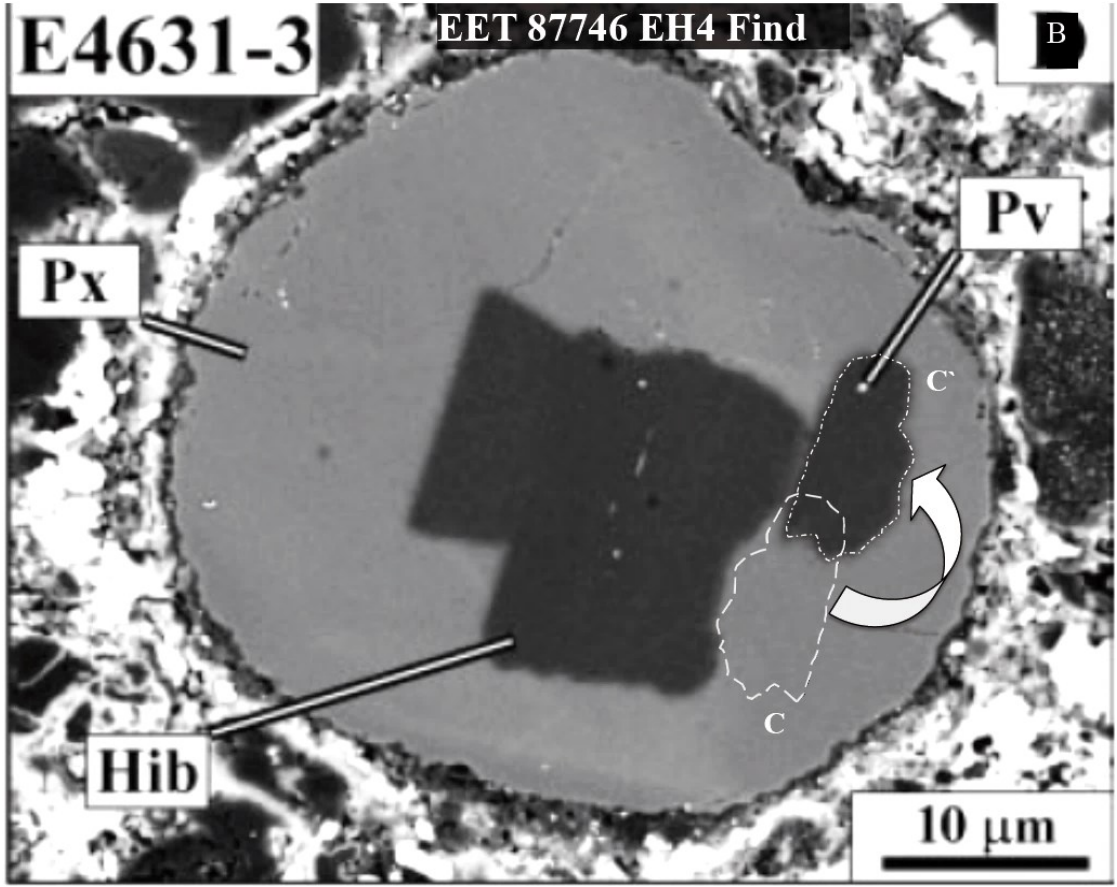

Table 1. Properties of Meteorites and features of hibonites-pyroxene spherules

| Meteorite | Sample Identifier | Petrologic Type | Fall/Find | Year found | Weathering grade | Total Mass of Meteorite | Dimensions HIXs (μm) | Hibonite lath (μm) | Texture of Pyroxene | Zoning in Pyroxene | $^{26}Al/^{27}Al \pm 2\sigma$ $(10^{-6})$ | F ($\delta^{25}$Mg ‰) in Pyroxene | REE pattern hibonite | Reference |
|---|---|---|---|---|---|---|---|---|---|---|---|---|---|---|
| ALHA77307 | RM-1 | CO3.03 | Find | 1977 | A | 181.3 g | 120×120 | 30×15 | Crystalline | no | 4.8±4.7 | 10 | nd | 8 |
| Colony | SP1 | CO3.1 | Find | 1975 | - | 3.91 kg | 170×170 | 70×25 | Devitrified | no | UR | - | II | 5 |
| Y-791717 | 17-6 | CO3.3 | Find | 1979 | - | 25.32 kg | 120×120 | 40×5 | Crystalline | yes | 200-300* | -6.2 | II | 3,6 |
| ALH 82101 | SP15 | CO3.4 | Find | 1982 | - | 29.1 g | 140×140 | 60×50 | Devitrified | no | 4.4±3.4 | - | II | 5 |
| Lance | 3413-1/31 | CO3.5 | Fall | 1872 | n/a | 51.7 kg | 50×50 | 20×10 | Glassy | no | UR | 8.2 | II | 1,4 |
| Murchison | 7-228 | CM2 | Fall | 1969 | n/a | 100 kg | 120×120 | 40×15 | Glassy | no | 17±7 | 9.2 | Frac, Enrich LREE | 4 |
| Murchison | 7-753 | CM2 | Fall | 1969 | n/a | 100 kg | 70×68 | 50×68 | Glassy | no | UR | -3.2 | Unusual | 4 |
| Murray | MY92S3 | CM2 | Fall | 1950 | n/a | 12.6 kg | 70×60 | 25×10 | Crystalline | yes | UR | -3.4 | II? | 6 |
| ALH 85085 | F-23 | CH3 | Find | 1985 | A/B | 11.9 g | 11×11 | 5×1.5 | Glassy | no | nd | nd | nd | 2 |
| ALH 85085 | - | CH3 | Find | 1985 | A/B | 11.9 g | 155×140 | 7×1.5 | Glassy | no | nd | nd | nd | 2 |
| EET 87746 | 4631-3 | EH4 | Find | 1987 | C | 142.3 g | 40×40 | 15×10 | Crystalline | ? | 2±5 | -5.3 | nd | 7 |

*In glass and no excess in hibonite [1] Kurat, 1975; [2] Grossman et al. 1988; [3]Tomoeka et al. 1992; [4] Ireland et al. 1991; [5] Russell et al. 1998; [6] Simon et al. 1998; [7]Guan et al. 2000; [8] present study. UR : unresolved excesses; nd: not determined

Table 2a. Electron microprobe analyses of hibonite

|  | 1 | 2 | 3 | 4 | 5 | 6 |
|---|---|---|---|---|---|---|
| $SiO_2$ | 0.04 | 0.04 | 0.04 | 0.04 | 0.04 | 0.06 |
| MgO | 0.82 | 0.77 | 0.77 | 0.80 | 0.85 | 0.83 |
| $Al_2O_3$ | 89.34 | 89.59 | 89.27 | 88.78 | 88.48 | 87.68 |
| CaO | 8.49 | 8.55 | 8.49 | 8.52 | 8.50 | 8.62 |
| $TiO_2^{tot}$ | 1.86 | 1.78 | 1.82 | 1.92 | 1.92 | 1.79 |
| FeO | 0.14 | 0.12 | 0.16 | 0.14 | 0.14 | 0.16 |
| Total | 100.69 | 100.84 | 100.55 | 100.20 | 99.93 | 99.14 |
| Cations per 19 oxygen anions | | | | | | |
| Si | 0.005 | 0.004 | 0.004 | 0.005 | 0.005 | 0.007 |
| Mg | 0.135 | 0.128 | 0.127 | 0.133 | 0.142 | 0.139 |
| Al | 11.680 | 11.693 | 11.686 | 11.669 | 11.661 | 11.653 |
| Ca | 1.009 | 1.015 | 1.010 | 1.018 | 1.019 | 1.041 |
| Ti | 0.155 | 0.148 | 0.152 | 0.161 | 0.161 | 0.152 |
| Fe | 0.013 | 0.011 | 0.015 | 0.013 | 0.013 | 0.015 |
| Total | 12.996 | 12.998 | 12.994 | 12.997 | 13.000 | 13.007 |

Table 2b. Electron microprobe analyses of pyroxene

|  | 1 | 2 | 3 | 4 | 5 | 6 |
|---|---|---|---|---|---|---|
| $SiO_2$ | 35.62 | 36.47 | 38.37 | 46.57 | 47.26 | 46.79 |
| MgO | 5.93 | 6.38 | 7.61 | 13.01 | 13.77 | 13.45 |
| $Al_2O_3$ | 30.12 | 29.41 | 26.15 | 14.45 | 13.67 | 14.05 |
| CaO | 25.51 | 25.58 | 25.54 | 25.77 | 25.35 | 25.31 |
| $TiO_2^{tot}$ | 1.80 | 1.41 | 1.53 | 0.03 | 0.03 | 0.03 |
| $Cr_2O_3$ | 0.06 | 0.04 | 0.05 | 0.07 | 0.09 | 0.07 |
| FeO | 0.18 | 0.20 | 0.23 | 0.32 | 0.35 | 0.35 |
| Total | 99.21 | 99.49 | 99.48 | 100.22 | 100.52 | 100.05 |
| Cations per 6 oxygen anions | | | | | | |
| Si | 1.306 | 1.332 | 1.401 | 1.681 | 1.699 | 1.690 |
| Mg | 0.324 | 0.347 | 0.414 | 0.700 | 0.738 | 0.724 |
| Al | 1.302 | 1.266 | 1.126 | 0.615 | 0.579 | 0.598 |
| Ca | 1.003 | 1.001 | 0.999 | 0.997 | 0.976 | 0.980 |
| Ti | 0.050 | 0.039 | 0.042 | 0.001 | 0.001 | 0.001 |
| Cr | 0.002 | 0.001 | 0.001 | 0.002 | 0.002 | 0.002 |
| Fe | 0.006 | 0.006 | 0.007 | 0.010 | 0.010 | 0.011 |
| Total | 3.991 | 3.992 | 3.991 | 4.006 | 4.006 | 4.005 |

Table 3. $^{26}$Al-$^{26}$Mg isotope data of hibonite-pyroxene spherule (RM-1) in ALHA77307 (CO3.03).

| Mineral | $^{27}$Al/$^{24}$Mg | 2σ | Δ$^{26}$Mg | 2σ | δ$^{25}$Mg | 2σ |
|---|---|---|---|---|---|---|
| **Hibonite-1** | 94.9 | 1.9 | 10.4 | 10.9 | 9.0 | 5.1 |
| **Hibonite -2** | 92.1 | 1.9 | 5.2 | 4.3 | 11.7 | 2.0 |
| **Hibonite -3** | 101.8 | 1.9 | 5.4 | 4.7 | 12.3 | 2.1 |
| **Pyroxene-1** | 3.4 | 0.1 | 1.8 | 1.8 | 9.8 | 0.8 |
| **Pyroxene-1** | 3.5 | 0.1 | 2.6 | 1.8 | 9.4 | 0.8 |
| **Pyroxene-1** | 4.0 | 0.1 | 3.2 | 2.1 | 9.9 | 0.9 |
| **Pyroxene-1** | 3.5 | 0.1 | 2.7 | 1.9 | 9.4 | 0.8 |

Table 4. Bulk compositions of Hibonite-Pyroxene Spherules

| Element Oxide | Allan Hills 77307 RM-1 | | | | Colony SP1 | Yamato 791717-6 | ALH 82101 SP15 | Lance 3413-1/31 | Murray MYSM3 | Murchison Mur 7-228 | Murchison Mur 7-753 |
|---|---|---|---|---|---|---|---|---|---|---|---|
| | Hibonite | Pyroxene | Pyroxene* | Bulk Calc. | Bulk | Bulk | Bulk | Bulk | Bulk | Bulk | Bulk |
| $SiO_2$ | 0.0 | 37.0 | 46.2 | 32.9 | 26.7 | 29.3 | 27.2 | 33.1 | 31.3 | 39.4 | 27.3 |
| MgO | 0.9 | 6.7 | 12.8 | 6.4 | 5.1 | 5.1 | 5.1 | 6.3 | 6.1 | 7.2 | 5.3 |
| $Al_2O_3$ | 88.5 | 29.0 | 14.6 | 35.4 | 44.9 | 39.3 | 44.0 | 35.1 | 37.4 | 28.9 | 47.0 |
| CaO | 8.5 | 25.4 | 25.7 | 23.3 | 20.1 | 23.1 | 21.3 | 23.8 | 23.5 | 22.6 | 17.9 |
| $TiO_2$ | 1.9 | 1.2 | 0.0 | 1.3 | 2.2 | 3.2 | 3.3 | 1.7 | 1.7 | 1.8 | 2.6 |
| Sum | 99.8 | 99.3 | 99.4 | 99.3 | 98.9 | 100.0 | 100.9 | 100.0 | 100.0 | 99.9 | 100.1 |

Data from relevant literature referred previously in table 1. [4] Ireland et al. 1991; [5] Russell et al. 1998; [6] Simon et al. 1998; * Indicates average composition of (n = 10) pyroxene near the sliver zone.